\documentstyle[12pt,epsf]{article} 

\makeatletter
\def\lesssim{\mathrel{\mathpalette\vereq<}}
\def\vereq#1#2{\lower3pt\vbox{\baselineskip1.5pt \lineskip1.5pt
\ialign{$\m@th#1\hfill##\hfil$\crcr#2\crcr\sim\crcr}}}

\makeatother

\begin{document}

\title{
{\normalsize E-print hep-ph/9812425 \hfill Preprint YARU-HE-98/09} \\[5mm]
{\Large\bf Field-induced axion decay $a \to e^+ e^-$ in plasma}} 

\author{\Large N.V.~Mikheev, A.Ya.~Parkhomenko and L.A.~Vassilevskaya \\ 
       {\it Yaroslavl State (Demidov) University,} \\ 
       {\it Sovietskaya 14, Yaroslavl 150000, Russia}} 

\date{}

\maketitle

\begin{abstract}
The axion decay $a \to e^+ e^-$ is investigated in the presence 
of a plasma and an external magnetic field. 
The results demonstrate a strong catalyzing influence of medium. 
The axion lifetime in the magnetic field of order $10^{15}$~G and 
at the temperature of order 10~MeV is reduced to $10^4$~s. 
\end{abstract}

\section{Introduction}

The Peccei-Quinn (PQ) symmetry $U_{PQ} (1)$~\cite{Peccei77}, with 
its accompanying axion~\cite{WW}, continues to be the most attractive 
solution to the strong CP problem in QCD. At present axions are of 
great interest not only in theoretical aspects of elementary particle 
physics, but in some astrophysical and cosmological applications as 
well~\cite{Turner,Raffelt90,Raffelt-book}. Although the original axion 
is excluded experimentally, modified PQ models with very light and  
weakly coupled ``invisible'' axions are still 
tenable~\cite{Raffelt-book,Peccei96}. Invisible axion models are 
classified into two types depending on whether or not they have 
a direct coupling to leptons. In KSVZ model~\cite{KSVZ} axions have only 
induced coupling to leptons. In DFSZ model~\cite{DFSZ} axions couple to 
leptons at tree level.

At present axions are strongly constrained by astrophysical and 
cosmological considerations which leave a rather narrow 
window~\cite{Raffelt-castle97,Raffelt-school97,Raffelt-flor98}: 
\begin{equation} 
10^{-5} \, \mbox{eV} \lesssim m_a \lesssim 10^{-2} \, \mbox{eV} 
\label{eq:mass} 
\end{equation} 
in which they can exist and provide a significant fraction or all of  
the cosmic dark matter. The extremely small value of the axion mass 
$m_a$ and, consequently, a gigantic axion lifetime in vacuum: 
\begin{equation} 
\tau^{(0)} \sim 6.3 \cdot 10^{42} \, \mbox{s} \, 
\left ( {10^{-2} \, \mbox{eV} \over m_a} \right )^6 \; 
\left ( {E_a \over 1 \, \mbox{MeV}} \right ) 
\label{eq:tau-vac} 
\end{equation} 
does not leave any hope to observe $a \to \gamma \gamma$ decay in 
laboratory conditions. 

It is known that an external electromagnetic field can affect 
substantially decays of very light, weakly interacting particles. 
The examples of the strong catalyzing influence of the field on 
processes allowed in vacuum are radiative decays of the massive 
neutrino $\nu \to \nu' \gamma$~\cite{GMV} and the axion 
$a \to \gamma \gamma$~\cite{MV-agg}.
One of the results obtained in~\cite{GMV,MV-agg} is that the external 
field removes the main suppression caused by the smallness of the 
decaying particle mass. To illustrate this for $a \to \gamma \gamma$ 
decay~\cite{MV-agg} we give the comparison of the decay probability 
$W^{(F)}$ induced by the field with the vacuum one $W^{(0)}$: 
\begin{eqnarray} 
{W^{(F)} \over W^{(0)}}  \simeq 10^{33} \; 
\left ( {10^{-2} \, \mbox{eV} \over m_a} \right )^4 \; P(\chi), 
\nonumber 
\end{eqnarray} 
where $P(\chi)$ is some function of the field dynamical parameter. 

Another special feature of the external field influence is that the  
field can open novel channels forbidden in vacuum. In particular, 
the axion decay into electron-positron pair $a \to e^+ e^-$ is opened 
due to a specific kinematics of charged particles in the field. 
We have studied this decay in two classes of invisible axion 
models~\cite{MV-aff,MOV-agee}. It was shown~\cite{MV-aff} that the 
axion lifetime in DFSZ model is reduced to $10^5$~s in the case of  
the decaying axion energy $\sim 1$~MeV in the magnetic field of 
strength $\sim 10^{15}$~G while for KSVZ axions the lifetime becomes 
even of order seconds~\cite{MOV-agee} for the above mentioned parameters.

However considering axions effects in astrophysical and cosmological
environments it is important to take into account not only the 
magnetic field influence, as it was done in~\cite{MV-aff,MOV-agee}, 
but the plasma one as well. The most physically realistic situation
is that when from both components of the active medium the plasma 
dominates. So, the magnetic field $B$ is relatively weak
\begin{eqnarray}
e B \ll \mu^2, T^2
\label{eq:cond}
\end{eqnarray}
(where $\mu$ and $T$ are the electron chemical potential and
temperature, respectively) and a great number of the Landau levels 
is excited. Whereas the condition $T^2 \gg e B$ is fulfilled, the 
magnetic filed is strong enough, $e B \gg m_e^2$, in comparison with
the known Schwinger value $B_e = m^2_e/e \simeq 4.41 \cdot 10^{13}$~G. 
Possible mechanisms to generate fields as strong as 
$B \sim 10^{15} - 10^{17}$~G in astrophysical 
objects~\cite{magnetar,toroidal} and in the 
early Universe~\cite{cosmol} are widely discussed. 

In this paper we investigate the field-induced axion decay into 
electron-positron pair $a \to e^+ e^-$ in the plasma. 

\section{Matrix Element}

\begin{figure}[tb] 
%
\centerline{\epsfxsize=0.9\textwidth \epsffile[125 520 465 715]{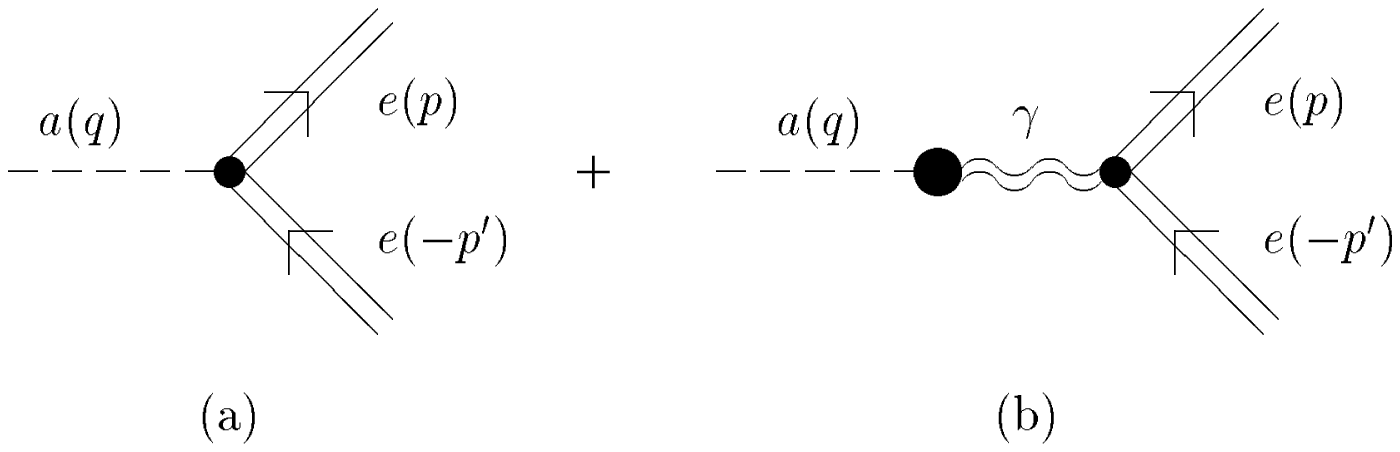}} 
\caption{} 
\label{fig:agee-sum} 
\end{figure}

The axion decay $a \to e^+ e^- $ is described by two diagrams in 
Fig.~\ref{fig:agee-sum} 
where solid double lines imply the influence of the magnetic field 
on the electron wave functions and undulating double lines imply 
the influence of medium on the photon propagator. The diagram~(a) 
describes this process in the model~\cite{DFSZ} with a direct 
axion-fermion coupling: 
\begin{equation} 
{\cal L}_{af} =  - i g_{af}\;(\bar f \gamma_5 f) \, a, 
\label{eq:L-aff} 
\end{equation} 
where $g_{af} = C_f m_f / f_a$ is a dimensionless coupling constant; 
$f_a$ is the Peccei-Quinn symmetry breaking scale;  
$C_f$ is a model-dependent factor;  
$m_f$ is the fermion mass (electron in our case); 
$f$ and $a$ are the fermion and axion fields, respectively. 

\begin{figure}[tb] 
%
\centerline{\epsfxsize=0.9\textwidth \epsffile[135 500 455 750]{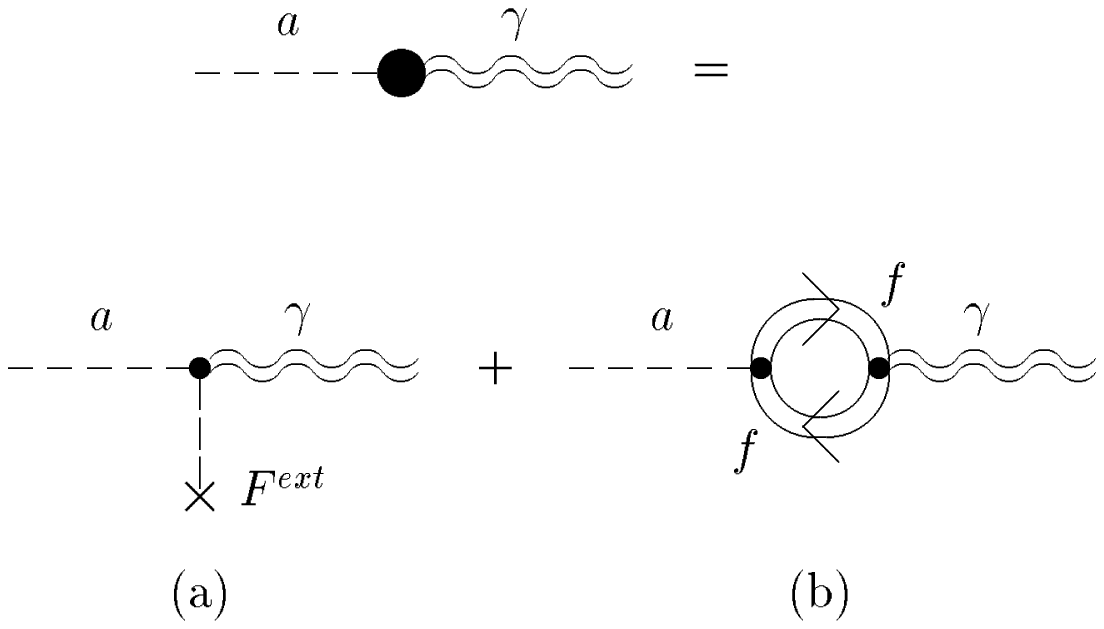}} 
\caption{} 
\label{fig:ag-eff} 
\end{figure}

The diagram~(b) describes the decay via a photon intermediate state. 
The effective axion-photon coupling can be presented in the form: 
\begin{eqnarray} 
{\cal L}_{a \gamma} = \bar g_{a\gamma} \, ( \partial_\mu A_\nu ) \, 
\tilde F_{\nu \mu} \, a , 
\label{eq:Lag}  
\end{eqnarray} 
where $A_\mu$ is the four potential of the quantized electromagnetic 
field, $\tilde F$ is the dual external field tensor; 
$\bar g_{a\gamma}$ is an effective coupling in the presence of the 
magnetic field with the dimension $(energy)^{-1}$~\cite{MRV}: 
\begin{eqnarray} 
\bar g_{a\gamma} =  g_{a \gamma} + \frac{\alpha}{\pi} \;  
\sum_f \frac{Q_f^2 g_{af}}{m_f} \; \left ( 1 - J \right ) . 
\label{eq:Gag1} 
\end{eqnarray} 
Here, $g_{a \gamma}$ corresponds to the well-known $a\gamma\gamma$ 
coupling in vacuum (diagram~(a) in Fig.~\ref{fig:ag-eff}) with a 
constant $g_{a \gamma} = \alpha \xi/ 2 \pi f_a $~\cite{Raffelt90}, 
where $\xi$ is the model-dependent parameter.
The second term in~(\ref{eq:Gag1}) is the field-induced contribution 
to the effective coupling $\bar g_{a\gamma}$ which comes from a 
diagram~(b) in Fig.~\ref{fig:ag-eff}:
\begin{eqnarray}
J & = & \left ( {4 \over \chi_f} \right )^{2/3} \, 
\int \limits_0^{\pi/2} \; 
f(\eta) \, \sin^{- 1/3} \phi \, d\phi, 
\nonumber \\ 
f(\eta) & = & i \,\int \limits_0^\infty du \, \exp 
\left \lbrace 
-i \, \left ( \eta u + {u^3 \over 3} \right ) 
\right \rbrace , 
\nonumber \\ 
\eta & = & \left ( {4 \over \chi_f \sin^2 \phi} \right )^{2/3} . 
\nonumber 
\end{eqnarray} 
Here, $f (\eta)$ is the Hardy-Stokes function and 
$\chi_f$ is the dynamic parameter: 
\begin{eqnarray} 
\chi_f^2 = \frac{e_f^2 (qFFq)}{m_f^6} \; , 
\label{eq:Chi} 
\end{eqnarray} 
where $e_f = e Q_f$, $e>0$ is the elementary charge, 
$Q_f$ is a relative electric charge of a loop fermion; 
$q = (E_a, {\bf q})$ is the four-momentum of the axion. 
As it was pointed in~\cite{MRV} this term gives a contribution comparable 
with $g_{a \gamma}$ for the fermions with $\chi_f \gg 1$ only.

The decay probability of particles in the magnetic field depends on 
invariant field parameters. In a general case of arbitrary values of 
particles energies and the magnetic field strength two independent 
field invariants $(e^2 (FF))^{1/2}$ and $(e^2 (qFFq))^{1/3}$ exist.  
Therefore the decay probability is a function of these parameters. 
In the case of a relatively weak magnetic field, when a great number 
of the Landau levels is excited, and ultrarelativistic particles the 
invariant $(e^2 (qFFq))^{1/3}$ occurs the largest one. So, the decay 
probability depends actually on the dynamic parameter 
$\chi_f$~(\ref{eq:Chi}) only. It means that we can perform calculations 
in a crossed field (${\bf E} \perp {\bf B}$, $E = B$) 
with zeroth field invariants $(FF)$ and $(F \tilde F)$. 

The matrix element of $a \to e^+ e^-$ decay corresponding 
to the diagrams in Fig.~\ref{fig:agee-sum} is a sum: 
\begin{equation} 
S = S^{(a)} + S^{(b)} , 
\label{eq:S1} 
\end{equation} 
where $S^{(a)}$ corresponds the diagram~(a) in Fig.~\ref{fig:agee-sum}:  
\begin{eqnarray} 
S^{(a)} = \frac{g_{ae}}{\sqrt{2 E_a V}} \int d^4 x\, 
\bar \psi (p,x) \,\gamma_5 \, \psi(-p',x) \, e^{-iqx}, 
\nonumber  
\end{eqnarray} 
and $S^{(b)}$ describes the contribution from the diagram~(b): 
\begin{eqnarray} 
S^{(b)} & = &\frac{g_{a\gamma}}{\sqrt{2 E_a V}}\int d^4 x\, 
\bar \psi (p,x) \, (\gamma h) \, \psi(-p',x) \, e^{-iqx}, 
\nonumber \\ 
h_\alpha & = & 
- i e (q \tilde F G (q))_\alpha = 
- i e q_\mu \tilde F_{\mu\nu} G_{\nu\alpha}(q). 
\nonumber 
\end{eqnarray} 
Here, $\psi(p,x)$ is the exact solution of the Dirac equation 
in the external crossed field~\cite{BLP}; 
$p = (E, {\bf p})$ and $p'=(E', {\bf p}')$ are the quasi-momenta 
of final electron and positron ($p^2 = {p'}^2 = m_e^2$); 
$(\gamma h) = \gamma_\mu h_\mu$, 
$\gamma_\mu$ are the Dirac $\gamma$-matrices. 
The condition of the relative weakness of the magnetic field, 
$e B \ll T^2, \mu^2$, means that from both components of the active 
medium the plasma determines basically the properties of the photon 
propagator $G_{\alpha \beta}$. The propagator can be presented as a 
sum of transverse and longitudinal parts:
\begin{eqnarray}
G_{\alpha \beta}(q) & = & - i \left ( 
  \frac{{\cal P}^{(T)}_{\alpha\beta}}{q^2 - \Pi^{(T)}} 
+ \frac{{\cal P}^{(L)}_{\alpha\beta}}{q^2 - \Pi^{(L)}} 
\right ),
\label{eq:G} \\
{\cal P}^{(T)}_{\alpha\beta} & = & - \sum^2_{\lambda = 1} 
t_\alpha^{(\lambda)} \; t_\beta^{(\lambda)},
\qquad 
{\cal P}^{(L)}_{\alpha\beta} = - \, l_\alpha \; l_\beta . 
\nonumber
\end{eqnarray}
Here, $\Pi^{(T)}$ and $\Pi^{(L)}$ are the transverse and longitudinal 
eigenvalues of the polarization operator;
$t_\alpha^{(\lambda)}$ ($\lambda=1,2$) and $l_\alpha$ 
denote the transverse and longitudinal photon polarization vectors:
\begin{eqnarray}
t^{(1)}_\alpha & = & \frac{(qF)_\alpha}{\sqrt{(qFFq)}}, \quad
t^{(2)}_\alpha = \frac{\varepsilon_{\alpha \beta \mu \nu} 
t^{(1)}_\beta q_\mu u_\nu}{\sqrt{(uq)^2 - q^2}},
\nonumber \\
\ell_\alpha & = & \sqrt{\frac{q^2}{(uq)^2 - q^2}}\,
\left(u_\alpha - \frac{uq}{q^2}\,q_\alpha \right),
\nonumber
\end{eqnarray}
\noindent where $u_\alpha$ is the four-velocity of medium.

Being integrated over the variable $x$ the matrix 
element~(\ref{eq:S1}) is:
\begin{eqnarray}
S & = & 
\frac{\delta^{(2)} ({\bf Q}_\perp) \; \delta(k Q)} 
     {\sqrt{2 E_a V \cdot 2 E V \cdot 2 E' V}} \; 
\frac{(2 \pi)^4}{\pi u z} 
\label{eq:S2} \\ 
& \times & 
\bar U (p) \; \bigg [ \; 
g_{ae} \gamma_5 \left ( \Phi (\eta) + \frac{i e z^2}{2 m^2_e} 
(\gamma F \gamma) \; \Phi'(\eta) \right )
\nonumber \\ 
& + & 
g_{a \gamma} \bigg ( 
(\gamma h) \; \Phi (\eta) 
+ \frac{i e z^2 (\chi' - \chi)}{m_e^2 \chi_a} \; 
(\gamma F h) \; \Phi' (\eta) 
\nonumber \\ 
& - & \frac{e z^2}{m_e^2} \; \gamma_5  
(\gamma \tilde F h) \; \Phi' (\eta) 
+ \frac{m_e^2}{2 z^2}\; \frac{(\gamma k) (k h)}{(k p)(k p')} \; 
\eta \; \Phi (\eta) \bigg ) \bigg ] \; U (- p'),
\nonumber \\
u^2 & = & - \frac{e^2 a^2}{m_e^2}, \qquad  
z = \left (\frac{\chi_a}{2 \chi \chi'} \right )^{1/3},
\nonumber \\
\chi^2 & = & \frac{e^2 (p F F p)}{m_e^6}, \qquad 
{\chi'}^2 = \frac{e^2 (p' F F p')}{m_e^6}, \qquad 
\chi_a^2 = \frac{e^2 (q F F q)}{m_e^6} .  
\nonumber
\end{eqnarray}
Here, $F_{\mu\nu} = k_\mu a_\nu - k_\nu a_\mu$ ($k^2 = (ka) = 0$) 
is the crossed field tensor; $Q = q - p - p'$, ${\bf Q}_\perp$  
is the perpendicular to $\bf k$ component (${\bf Q}_\perp {\bf k}= 0$). 
The bispinor $U(p)$, which is normalized by the condition 
$\bar U U = 2 m_e$, satisfies the Dirac equation for the free electron 
$((\gamma p) - m_e) U(p) = 0$. 
Finally, $\Phi (\eta)$ is the Airy function: 
\begin{eqnarray}
\Phi (\eta) & = &  \int\limits_0^\infty d t \cos \left ( \eta t + 
{t^3\over 3} \right ),  
\label{eq:Ai} \\  
\eta & = & z^2 \; (1 + \tau^2), \qquad 
\tau = - \, \frac{e (p \tilde F q)}{m_e^4 \chi_a}, 
\nonumber 
\end{eqnarray}
and $\Phi' (\eta) = d \Phi (\eta)/ d \eta$.  

\section{Decay Probability} 

After integration over the phase space of the $e^+ e^-$ pair 
the decay probability is: 
\begin{eqnarray} 
W & \simeq & \frac{g_{ae}^2 3^{5/3}}{16 \pi^3} \; 
\Gamma^4 (2/3) \; \frac{m_e^2 \chi_a^{2/3}}{E_a} \; 
\rho_1(E_a,T,\mu) 
\label{eq:prob-tot} \\ 
& + & \frac{\bar g_{a\gamma}^2 (e B)^2}{36 \pi} \,  
\frac{E_a^3 \, \cos^2 \theta} 
     {(E_a^2 - {\cal E}^2)^2 + \gamma^2 {\cal E}^4} \, 
\rho_2(E_a,T,\mu), 
\nonumber \\ 
\rho_1 & = & \frac{2 \pi}{3 \sqrt 3 \; \Gamma^3 (2/3)} \; 
\int\limits_0^1 \frac{dx}{x^{1/3} (1 - x)^{1/3}} \, 
(1 - n) \, (1 - \bar n) ,  
\nonumber \\ 
\rho_2 & = & 6 \int\limits_0^1 dx \, x (1 - x) \, 
(1 - n) \, (1 - \bar n) ,  
\nonumber \\ 
n & = & \left ( \exp \frac{x E_a - \mu}{T} + 1 \right )^{-1} , 
\qquad
\bar n =  
\left ( \exp \frac{(1 - x) E_a + \mu}{T} + 1 \right )^{-1} , 
\nonumber 
\end{eqnarray} 
where the variable $x = E/E_a$ is the relative electron energy; 
$n$ and $\bar n$ are the Fermi-Dirac distributions of electrons 
and positrons, respectively. The functions $\rho_{1,2}$ have a 
meaning of the average values of suppressing statistical factors 
and are, in general case, inside the interval $0 < \rho_{1,2} < 1$. 

The second term in~(\ref{eq:prob-tot}) describes the contribution 
of the longitudinal plasmon intermediate state only and has a 
resonant character at a particular energy of the decaying axion 
$E_a \sim {\cal E}$. This is due to the fact that the axion and the 
longitudinal plasmon dispersion relations always cross for a certain 
wave-number $k = {\cal E}$. The contribution of the transverse plasmon 
to the decay probability of $a \to  e^+ e^-$ in the ultrarelativistic 
case is negligibly small. The dimensionless resonance width $\gamma$ 
in Eq.~(\ref{eq:prob-tot}) is: 
\begin{equation}
\gamma = \frac{{\cal E} \Gamma_L({\cal E})}{q^2 Z_L},
\label{eq:gamma1}
\end{equation}
where $\Gamma_L ({\cal E})$ is the total width of the 
longitudinal plasmon; $Z_L$ is the renormalization factor 
of the longitudinal plasmon wave-function: 
\begin{equation} 
Z_L^{-1} = 1 - \frac{\partial \, \Pi^{(L)}}{\partial \, q_0^2}. 
\label{eq:Gamma1} 
\end{equation} 
Note that without the external field the plasmon decay into neutrino 
pair is kinematically allowed only. In the presence of the strong 
magnetic field ($e B \gg \alpha^3 E^2$) the main contribution to the 
width $\Gamma_L ({\cal E})$ is determined by the process of the 
longitudinal plasmon absorption $\gamma_L e^- \to e^-$~\cite{MRV} 
which becomes possible in this kinematical region also. 

Below we give the expressions for ${\cal E}^2$ and $\gamma$  
in two cases: 
\newline 
i) degenerate plasma ($\mu \gg T$) 
\begin{eqnarray} 
{\cal E}^2 & \simeq & \frac{4 \alpha}{\pi} \, \mu^2 \, 
\left ( \ln \frac{2 \mu}{m_e} - 1 \right ) , 
\label{eq:degen} \\ 
\gamma & \simeq & \frac{2 \alpha}{3} \, \frac{\mu^2}{{\cal E}^2}, 
\nonumber 
\end{eqnarray} 
ii) nondegenerate hot plasma ($T \gg \mu$) 
\begin{eqnarray} 
{\cal E}^2 & \simeq & \frac{4 \pi \alpha}{3} \, T^2 \, 
\left ( \ln \frac{4 T}{m_e} - 0.647 \right ) . 
\label{eq:nondegen} \\ 
\gamma & \simeq & \frac{2 \pi^2 \alpha}{9} \, \frac{T^2}{{\cal E}^2}. 
\nonumber 
\end{eqnarray} 
Considering possible applications of the result to cosmology 
we estimate the axion lifetime in a hot plasma. Under the early 
Universe conditions the hot plasma is nondegenerate one and the 
medium parameters $\rho_{1,2}$ are inside the interval 
$1/4 < \rho_{1,2} < 1$. With ${\cal E}^2$ and $\gamma$~(\ref{eq:nondegen}) 
we obtain the following estimation for the axion lifetime 
(diagram~(b) in Fig.~\ref{fig:agee-sum}) in the resonance region:
\begin{eqnarray}
\tau (a \to \gamma_{pl} \to e^+ e^-) \simeq 
 6.1 \cdot 10^4 \, \mbox{s} \, 
\left ( \frac{10^{-10}}{\bar g_{a \gamma} \, \mbox{GeV}} \right )^2 \, 
\left ( \frac{T}{10 \, \mbox{MeV}} \right ) \,
\left ( \frac{10^{15} \, \mbox{G}}{B} \right )^2.
\label{eq:time-KSVZ} 
\end{eqnarray}
It is interesting to compare Eq.~(\ref{eq:time-KSVZ}) 
with the axion lifetime without taking into account 
the resonant contribution via plasmon 
\begin{eqnarray}
\tau (a \to e^+ e^-) \simeq 3.4 \cdot 10^6 \, \mbox{s} \, 
\left ( \frac{10^{-13}}{g_{a e}} \right )^2 \, 
\left ( \frac{T}{10 \, \mbox{MeV}} \right )^{1/3} \, 
\left ( \frac{10^{15} \, \mbox{G}}{B} \right )^{2/3}. 
\label{eq:time-DFSZ} 
\end{eqnarray} 

The expressions~(\ref{eq:time-KSVZ}) and~(\ref{eq:time-DFSZ}) 
demonstrate the strong catalyzing influence of medium, the plasma 
and the magnetic field, on the axion decay. Nevertheless, 
the axion lifetime is determined by the axion decay via the 
longitudinal plasmon in both invisible axion models~\cite{KSVZ,DFSZ}.

\section*{Acknowledgements}

This work was partially supported by INTAS under grant No.~96-0659 
and by the Russian Foundation for Basic Research 
under grant No.~98-02-16694.
The work of N.V.~Mikheev was supported under grant No.~d98-181
by International Soros Science Education Program.

\newpage

%
%
%
%

\newpage

%
%
%
%

\end{document}